\documentclass[12pt]{article}
\usepackage{a4}
\usepackage{amsfonts,amssymb,amsmath,amsthm}
\usepackage{graphicx}

\newcommand{\Dir}{{\widehat D}}

\begin{document}
\title{Quantum noncommutative gravity in two dimensions}

\author{ Dmitri V. Vassilevich\thanks{Also at V.~A.~Fock Institute of
Physics, St.Petersburg University, Russia; email:
Dmitri.Vassilevich@itp.uni-leipzig.de}\\
{\it Institut f\"{u}r Theoretische Physik, Universit\"{a}t
Leipzig,}\\
{\it D-04109 Leipzig, Germany } }

\maketitle
\begin{abstract}
We study quantisation of noncommutative gravity theories in two
dimensions (with noncommutativity defined by the Moyal star
product). We show that in the case of noncommutative 
Jackiw-Teitelboim gravity the path integral over gravitational
degrees of freedom can be performed exactly even in the presence
of a matter field. In the matter sector, we study possible choices
of the operators describing quantum fluctuations and define their basic
properties (e.g., the Lichnerowicz formula). Then we evaluate
two leading terms in the heat kernel expansion, calculate
the conformal anomaly and the Polyakov action (as an expansion
in the conformal field).
\end{abstract}
\section{Introduction}
Over recent years noncommutative field theory has developed into
a mature discipline (see reviews \cite{ncreviews}). However, 
formulation of a satisfactory
noncommutative counterpart of quantum gravity still remains an
open problem. There exist several approaches to noncommutative
gravity. One of them studies deformations of geometrical structures
(as, e.g.,  differential structures and exterior algebras). An overview
of this approach with the emphasis on two-dimensional models
and further references can be found in \cite{MaBur}. Although this
approach is very efficient for finding deformations of particular
geometries, it does not refer to any action functional. Therefore, it
is unclear how one could proceed with quantisation of such models.

In noncommutative theories the action functional can be constructed
by using the spectral action principle \cite{Chamseddine:1996zu}
which relates the action to the heat trace asymptotics of a 
suitable Dirac type operator. However, in many interesting
cases (as, e.g., gravity on the Moyal plane) such an operator is not
known so far (cf. \cite{Chamseddine:2003cw,Gayral:2004ww}).

In this paper we are interested in gravity theories where
noncommutativity appears, roughly speaking, due to the presence of
the Moyal star product (see eq.\ (\ref{starprod}) below).
A very fruitful approach to such theories is based on the gauge
theory formulation of general relativity. In the noncommutative
case the Lorentz group does not close and one has to deal with
an extended gauge symmetry \cite{Chamseddine:2000zu,Moffat:2000gr}. 
Here we are interested in two-dimensional gravity. Therefore, the 2D
model constructed in \cite{Cacciatori:2002ib} is of particular 
importance for us. This model shares some similarities with the
noncommutative 3D Chern-Simons gravity \cite{Banados:2001xw}.
We also like to mention 
four-dimensional gauge gravity models 
\cite{Cardella:2002pb,Obregon} and perturbative calculations in the
Einstein gravity \cite{Moffat,Quevedo} and of graviton scattering
on a D-brane \cite{Ardalan:2002qt}.

Dilaton gravities in two dimensions (see \cite{Grumiller:2002nm}
for a recent review)
have always been a good testing
ground for various theoretical ideas and methods of classical and
quantum general relativity. In particular, it was demonstrated
\cite{Kummer:1996hy} that in many cases the path integral for
such models can be calculated exactly. 
We would like to check whether this property can be found in
the noncommutative case as well.
So far, the Jackiw-Teitelboim (JT) model \cite{JT} is the only
two-dimensional dilaton gravity model which has a noncommutative
counterpart \cite{Cacciatori:2002ib}. Therefore, our study of the
path integral is restricted to the noncommutative JT gravity.
We show that in the temporal gauge, which tremendously simplifies
the analysis, one can indeed perform the path integration over
all gravity variables exactly and nonperturbatively (provided
quantum matter interactions with gravity satisfy some mild restrictions).
As a result, the full effective action becomes just a sum of
the classical action in the gravity sector and of an effective
action for the mater field calculated as if the 
gravity fields were a fixed background.

In the second part of the work we deal with the matter effective
action (so that the restriction to the JT model becomes inessential).
We use another important property of two-dimensional theories
which we would like to keep in the noncommutative case. Quantum
effective action for a matter field minimally coupled to gravity
is uniquely defined by the conformal anomaly and, therefore,
can be calculated exactly (giving the famous Polyakov action
\cite{Polyakov:1981rd}). In the noncommutative case the order in which
we multiply fields becomes essential. Therefore, even classical
analysis of the coupling of matter fields to gravity becomes
very complicated on the combinatorial side. To reduce this
complexity we work in the conformal gauge. The restriction to the
conformal gauge does not allow us to analyse symmetries of the
model, so that this problem is postponed. However, we show that in
the conformal gauge the requirements of hermiticity and of proper 
commutative limit are strong enough fix the fluctuation operators
for spinors and scalars almost uniquely. As a byproduct we derive
a Moyal extension of the Lichnerowicz formula. In the scalar case
we derive then the heat kernel expansion, calculate 
the conformal anomaly, and integrate it to obtain a noncommutative
version of the Polyakov action. Explicit formulae are given as an
expansion in the conformal field.

This paper is organised as follows. In the next section we review
some basic properties of the noncommutative JT gravity 
\cite{Cacciatori:2002ib}. Section \ref{spath} is devoted to 
the path integral over the gravitational degrees of freedom.
Effective action for the matter fields is studied in section
\ref{scona}. Section \ref{scon} contains a short overview
of solved and unsolved problems. Some useful formulae can be
found in Appendix \ref{appnot} and \ref{appsym}.
\section{Noncommutative Jackiw-Teitelboim gravity}\label{sJT}
The Moyal product of two functions $f$ and $g$ on $\mathbb{R}^2$ can
be defined by the equation
\begin{equation}
f\star g = f(x) \exp \left( \frac i2 \, \theta^{\mu\nu}
\overleftarrow{\partial}_\mu \overrightarrow{\partial}_\nu \right)
g(x) \,.\label{starprod}
\end{equation}
$\theta$ is a constant antisymmetric matrix.
In this form the star product
has to be applied to plane waves and then extended
to all (square integrable) functions by means of the Fourier series.
This product is known for a long time in the operator theory (cf.
\cite{BerezinShubin}) and in deformations of symplectic manifolds
\cite{Bayen:1977ha}.

A noncommutative deformation of the Jackiw-Teitelboim model has
been constructed in \cite{Cacciatori:2002ib}. It has been identified
with a $U(1,1)$ gauge theory on noncommutative $\mathbb{R}^2$
with the action
\begin{equation}
S=\int \mathrm{tr} \left( \Phi \star F \right) \,,\label{gact}
\end{equation}
where both fields $\Phi$ and $F$ take values in the Lie algebra
$u(1,1)$ of $U(1,1)$. The field $\Phi$ is a space-time scalar,
and $F$ is a two-form field strength with the components
\begin{equation}
F_{\mu\nu}=\partial_\mu A_\nu -\partial_\nu A_\mu 
+A_\mu \star A_\nu -A_\nu \star A_\mu \,.\label{Fmn} 
\end{equation}

The action (\ref{gact}) is invariant under usual (noncommutative)
$U_\star (1,1)$ gauge transformations
\begin{equation}
A\to g_\star^{-1} \star A \star g_\star +g_\star^{-1}\star d g_\star,\qquad
\Phi \to g_\star^{-1} \star \Phi \star g_\star \,.
\label{gautra}
\end{equation}
Next one expands $A$ and $\Phi$ over a basis $\tau_i$
in the defining 2-dimensional representation of $u(1,1)$: 
$A=\tau_i A^i$, $\Phi=\tau_i \Phi^i$. Precise form of this
basis can be found in \cite{Cacciatori:2002ib}. Then one introduces
new fields according to the equations
\begin{equation}
\Phi^i=(l\phi^a,\phi,\psi),\qquad A_\mu^i=(e_\mu^a l^{-1},\omega_\mu,b_\mu)\,,
\label{newfields}
\end{equation}
where $a=0,1$, and the scale $l$ is related to the cosmological
constant $\Lambda=-1/l^2$. $e_\mu^a$ plays the role of the zweibein,
${\varepsilon^a}_b\omega_\mu + i\delta^a_bb_\mu$ is identified
with an $so(1,1)\oplus u(1)$ connection. In terms of these new fields
the action (\ref{gact}) reads
\begin{equation}
S=\frac 14 \int d^2x\, \varepsilon^{\mu\nu} \left[ \phi_{ab} \star
\left( R_{\mu\nu}^{ab} -2\Lambda e_\mu^a \star e_\nu^b \right) -
2\phi_a \star T_{\mu\nu}^a \right] \label{actJT}
\end{equation}
with the curvature tensor
\begin{eqnarray}
&& R_{\mu\nu}^{ab}=\varepsilon^{ab} \left( \partial_\mu \omega_\nu
-\partial_\nu \omega_\mu +\frac i2 [\omega_\mu,b_\nu] +
\frac i2 [b_\mu,\omega_\nu] \right)\nonumber \\
&&\qquad +\eta^{ab} \left( i\partial_\mu b_\nu - i\partial_\nu b_\mu
+\frac 12 [\omega_\mu ,\omega_\nu] -\frac 12 [b_\mu,b_\nu ]\right)
\label{Rtensor}
\end{eqnarray}
and with the noncommutative torsion
\begin{eqnarray}
&&T_{\mu\nu}^a=\partial_\mu e^a_\nu -\partial_\nu e_\mu^a 
+\frac 12 {\varepsilon^a}_b \left( \{ \omega_\mu,e_\nu^b \}
-\{ \omega_\nu ,e^b_\mu \} \right) \nonumber \\
&&\qquad\qquad\qquad\qquad\ +\frac i2 \left( [b_\mu,e_\nu^a] -
[b_\nu,e_\mu^a] \right)\,. \label{torsion} 
\end{eqnarray}
The fields $\phi$ and $\psi$ are combined into
\begin{equation}
\phi_{ab}:=\phi \varepsilon_{ab} -i\eta_{ab} \psi \,.\label{pab}
\end{equation}
All commutators (denoted by square brackets) and anticommutators
(denoted by curved brackets) are calculated with the Moyal star product.
Further conventions and notations can be found in Appendix \ref{appnot}.

The fields $\Phi^i$ play the role of the Lagrange multipliers. $\phi_{ab}$
are responsible for a two-dimensional noncommutative version of the
Einstein equations. Variation of (\ref{actJT}) with respect to
$\phi_a$ gives the torsion constraint
\begin{equation}
\varepsilon^{\mu\nu}T_{\mu\nu}^a=0 \,.\label{tors}
\end{equation}

The gauge transformations (\ref{gautra}) can be rewritten in terms of
the component fields (\ref{newfields}) (see Appendix \ref{appsym}).
The four-parameter symmetry group contains Lorentz boosts, translations
(which coincide with the diffeomorphisms on shell \cite{Cacciatori:2002ib}),
and additional $U(1)$ gauge transformations which are needed to close
the gauge group in the noncommutative case. Note that the noncommutativity
parameter $\theta$ is not changed under these transformations.

If $\theta=0$ the fields $b_\mu$ and $\psi$ decouple, 
and the dynamics of the rest of the fields is described by the
commutative JT model. 

All classical solutions of the noncommutative JT model have been found in 
\cite{Cacciatori:2002ib}. In the present paper we deal with
quantum theory only.
\section{Exact path integral}\label{spath}
It was demonstrated in \cite{Kummer:1996hy} that in the JT model
the path integral over the gravitational degrees of freedom
can be performed exactly even in the presence of matter fields.
Here we extend this result to the noncommutative case. An important
ingredient is a convenient gauge choice which simplifies
the calculations enormously. Note, that the technique we employ here
is very general. It has been used in other dilaton gravity models
\cite{Kummer:1997jj,Kummer:1998zs} and in two-dimensional
supergravities \cite{Bergamin:2004us}. As a practical application
of this approach we may mention calculations of loop corrections
to the specific heat of the  dilaton black hole \cite{Grumiller:2003mc}.

To analyse the path integral we have to fix a gauge first. The action
(\ref{actJT}) looks particularly simple in the ``temporal'' gauge:
\begin{equation}
e_0^+=0,\quad e_0^-=1,\quad \omega_0=0, \quad b_0=0\,. \label{tempgau}
\end{equation}
Residual gauge freedom can be treated exactly as in the commutative
case \cite{Kummer:1996hy,Kummer:1998zs}.

Let us introduce the notations:
\begin{eqnarray}
&&q^i=(e_1^+,e_1^-,\omega_1,b_1)\,,\nonumber\\
&&p^i=(\phi_+,\phi_-,\phi,\psi)\,,\label{qpq}\\
&&\bar q^i=(e_0^+,e_0^-,\omega_0,b_0)\,,\nonumber
\end{eqnarray}
so that the gauge conditions (\ref{tempgau}) read: $\bar q^i-a^i=0$,
where $a^i:=(0,1,0,0)$. 

Consider a set of the matter fields $\{ f^\alpha \}$, where $\alpha$
numbers different components and types of the matter. Spin, statistics or gauge
groups play no role in the considerations of this section. 
If there is an additional gauge symmetry, $\{ f^\alpha \}$ should include
corresponding ghosts.
The only restriction we impose is that these field interact 
with the ``gauge fields'' $(e_\mu^a,\omega_\mu, b_\mu)=(q,\bar q)$, but not
with the ``Lagrange multipliers'' $(\phi_a,\phi,\psi)$.
The generating functional for the Green functions can be represented
as a path integral,
\begin{eqnarray}
&&Z(j,J)=\int \mathcal{D} q\, \mathcal{D} \bar q\, \mathcal{D} p\, 
\mathcal{D} f^\alpha 
\prod \delta (\bar q^i - a^i)\, \mathcal{F}\,\nonumber\\ 
&&\qquad\qquad\qquad\quad \times
\exp \left( iS + i S_{\mathrm{m}} (q,\bar q; f^\alpha)+
i\int d^2x(j_kq^k+J_kp^k) \right),\label{matZ}
\end{eqnarray}
where $S_{\mathrm{m}} (q,\bar q; f^\alpha)$ is a classical action for
the matter fields,
$\mathcal{F}$ is the Faddeev-Popov determinant corresponding
to our gauge choice. $j=(j_+,j_-,j_3,j_4)$ and 
$J=(J^+,J^-,J^3,J^4)$ are external sources. 
One can also introduce sources and/or background
fields for the matter. 

After imposing the gauge conditions (\ref{tempgau}) the action
(\ref{actJT}) becomes
\begin{equation}
S_{\mathrm{g.f.}}=\int d^2x \left[ \phi \partial_0 \omega_1
-\psi \partial_0 b_1 + \Lambda \phi e_1^+ + \phi_a \partial_0 e_1^a
+\phi_- \omega_1 \right]\,. \label{Sgf}
\end{equation}
Since the Moyal product is closed, one can omit stars in integrals
of all expressions quadratic in fields (as (\ref{Sgf})) provided
the fields fall off sufficiently fast at infinity so that one can
integrate by parts. This property can be most easily seen by integrating
(\ref{starprod}). 

Now we have to calculate the Faddeev-Popov determinant $\mathcal{F}$.
The most advanced technique for construction of the ghost action is
based upon the BRST formalism. This formalism has not been yet
fully adapted to space-time noncommutative theories (although some
steps in this direction have been already done, cf. \cite{Barnich:2003wq}).
Fortunately, since the gauge algebra in our case is of the Yang-Mills
type we may use a somewhat simpler Faddeev-Popov prescription
\cite{Faddeev:fc}. Note, that in the commutative 2D gravities
the Faddeev-Popov
approach gives correct results even though the
gauge group has field-dependent structure functions (cf. Appendix A of
\cite{Kummer:1998zs}). Therefore, we have 
on the surface defined by the gauge (\ref{tempgau})
\begin{equation}
\mathcal{F}=\det \left( \delta \bar q^i/\delta \lambda_j \right) 
=\det (\partial_0)^4\,,\label{FP2}
\end{equation}
where $\lambda_j =(\alpha^a,\xi,\chi)$ are the gauge parameters
(cf. Appendix \ref{appsym}). 

Since the action (\ref{Sgf}) is quadratic in the field, and since
the Faddeev-Popov determinant (\ref{FP2}) is field-independent,
it is clear that the integration over $p$ and $q$
in (\ref{matZ}) becomes trivial.
There are no essential differences to the path integral
calculations done in the commutative case \cite{Kummer:1996hy}.

Let us define the effective action $W_{\mathrm{m}}$ for the matter
fields:
\begin{equation}
W_{\mathrm{m}} (q,\bar q)=\frac 1i \ln \int \mathcal{D} f^\alpha
e^{i S_{\mathrm{m}} (q,\bar q; f^\alpha)} \,.\label{Wm}
\end{equation}
We assume that the measure $\mathcal{D} f^\alpha$ does not depend
on $p^j$.
The action (\ref{Wm}), as it is written here, 
contains contributions of all matter
loops on a background defined by $q$ and $\bar q$. 
One can restrict $W_{\mathrm{m}}$ to a finite order of the loop expansion.
Then our final result (see eq.\ (\ref{WPQ3}) below) will be restricted 
accordingly.
By integrating next over the
``momentum'' variables $p^i$ one obtains the following functional
delta-functions:
\begin{eqnarray}
&\delta (\partial_0 e_1^+ +J^+) \,,\qquad
&\delta (\partial_0 e_1^- +\omega_1 + J^-) \,,\label{del2} \\
&\delta (\partial_0 \omega_1 +\Lambda e_1^+ +J^3)\,,\qquad
&\delta (\partial_0 b_1 -J^4) \,.\label{del4}
\end{eqnarray}

Because of these delta-functions, integrations over  $q^i$ 
can be also performed exactly. Next we define the mean fields
\begin{equation}
Q=\frac 1i \frac{\delta \ln Z}{\delta j} \,,\qquad
P=\frac 1i \frac{\delta \ln Z}{\delta J}\,.\label{QP}
\end{equation}
and peform the Legendre transform of $\in Z$. The calculations
go exactly the same way as in the commutative case.
We refer to \cite{Kummer:1996hy} for details. The effective action reads
\begin{equation}
W(P,Q)=\frac 1i \ln\, Z -\int d^2 x (PJ+Qj) 
=S_{\mathrm{g.f.}} (P,Q)+ W_{\mathrm{m}}(Q)
\,.\label{WPQ3}
\end{equation}
This result means that all loop corrections due to the gravity fields
disappear\footnote{The gravity part of the action appears in (\ref{WPQ3})
in the gauge-fixed form. A part of the equations of motion (constraints)
is lost and has to be restored by using the Ward identities 
\cite{Kummer:1998zs}.}
if the matter action does not depend on $\phi_a$, $\phi$
and $\psi$. Of course, such a strong statement is made possible by
particular simplicity of the JT model. In more complicated dilaton
gravities with matter one has to perform a perturbative expansion already
in the commutative case \cite{Kummer:1998zs}.

We have not discussed asymptotic conditions, boundary terms, and other
``global'' issues. Therefore,
we call this kind of statements ``local quantum triviality'' although
locality looses its meaning in noncommutative theories.
\section{Conformal anomaly}\label{scona}
One usually starts calculations of the conformal anomaly with specifying
a classical action for the matter fields. Such an action should of course
respect all symmetries of the matterless action (see Appendix \ref{appsym}). 
In the noncommutative
limit this action should coincide with a standard action for, say,
scalar fields.
No such action is known, at least to the present author. The main difficulty
is that the diffeomorphism transformations in the noncommutative JT gravity
are realised in a very nontrivial way \cite{Cacciatori:2002ib}. This
fact reflects known problems with constructing covariant coordinate
transformations in noncommutative gauge theories (cf. \cite{Jackiw:2001jb}).

Therefore, instead of looking for a classical action for the matter
fields, we shall look for an operator which may describe the one-loop
corrections in a particular gauge. 
This idea is inspired by an approach to gravity on
noncommutative spaces base on spectral triples \cite{strip}.  
We shall not, however, follow this approach too closely. Our aim is not
to construct geometry starting from a spectral triple, but rather
to find a meaningful operator starting with geometrical objects 
of the noncommutative JT gravity. 

\subsection{Conformal gauge}\label{scg}
In this section we work on a space of the Euclidean signature.
Most of the formulae derived above remain valid after the substitution
$\eta^{ab}={\mathrm{diag}}(+1,+1)$, $\varepsilon^{12}=\varepsilon_{12}=1$.
This signature is more convenient to study the heat kernel expansion
because, for example, one deals with absolutely convergent integrals.

We need several more simplifying assumptions. First of all,
we put
\begin{equation}
b_\mu =0 \label{bmu0}
\end{equation}
since geometric meaning of this field is somewhat obscure. 
Simple degrees of freedom counting arguments show that now
the connection $\omega_\mu$ can be expressed through $e_\mu^a$
by means of the torsion constraint (\ref{tors}). 
It is not possible however to solve (\ref{tors}) explicitly unless
we impose the conformal gauge condition:
\begin{equation}
e_\mu^a=e^{\rho}\delta^a_\mu \,.\label{congauge}
\end{equation}
Here $e^{\rho}$ is a star exponent,
\begin{equation}
e^{\rho}=1+\rho + \frac 12 \rho \star \rho +
\frac 16 \rho \star \rho \star \rho + \dots \label{starex}
\end{equation}

The condition (\ref{congauge}) simplifies considerably the combinatorics
of all subsequent calculations  since now all components 
of $e_\mu^a$ commute with each
other. The torsion constraint reads
\begin{equation}
2e^{-\rho}\star \left( \partial_\mu e^{\rho} \right)=
-e^{-\rho}\star \hat \omega_\mu \star e^{\rho} -\hat \omega_\mu \,,
\label{ctors}
\end{equation}
where
\begin{equation}
\hat \omega_\mu ={\varepsilon_\mu}^\nu \omega_\nu =
\eta_{\mu\rho}\varepsilon^{\rho\nu} \omega_\nu \,.\label{hato}
\end{equation}
Next we use the Baker-Campbell-Hausdorf (BCH) formula to derive:
\begin{eqnarray}
&&e^{-\rho}\star \left( \partial_\mu e^{\rho} \right)=
\sum_{n=1}^\infty \frac 1{n!} [\dots[[\partial_\mu \rho,\rho],\rho],\dots],
\label{BCHrho}\\
&&e^{-\rho}\star \hat \omega_\mu \star e^{\rho}=
\sum_{k=0} \frac 1{k!} [\dots [[\hat \omega_\mu, \rho],\rho],\dots].
\label{BCHom}
\end{eqnarray}
The $n$th term in (\ref{BCHrho}) contains $n-1$ commutators, while
the $k$th term in (\ref{BCHom}) contains $k$ commutators. One can easily prove
that
\begin{equation}
\hat \omega_\mu = \sum_{n=1}^\infty c_n 
[\dots[[\partial_\mu \rho,\rho],\rho],\dots]
\label{solom}
\end{equation}
(with $n-1$ commutators in the $n$th term), where $c_1=-1$ and the subsequent
coefficients are given by the recursion:
\begin{equation}
c_n=-\frac 1{n!} -\frac 12 \sum_{k=1}^{n-1} \frac{c_{n-k}}{k!} \,.\label{cn}
\end{equation}
In particular, all even-numbered coefficients vanish, $c_{2k}=0$, and
\begin{equation}
\hat \omega_\mu = -\partial_\mu \rho 
+ \frac 1{12} [[\partial_\mu \rho,\rho],\rho] + \mathcal{O}(\rho^5).
\label{expom}
\end{equation}
The expansion (\ref{solom}) is, by the construction, an expansion in
$\rho$. However, since it contain repeated commutators, it is also
an expansion in the noncommutativity parameter $\theta$.

Inspired by (\ref{actJT}) one can define the scalar curvature density
as
\begin{equation}
\mathcal{R}=\frac 12 \varepsilon^{\mu\nu}\varepsilon_{ab}R^{ab}_{\mu\nu}\,.
\label{scd}
\end{equation}
If $b=0$,
\begin{equation}
\mathcal{R}=2\varepsilon^{\mu\nu}\partial_\mu \omega_\nu =
2\partial_\mu \hat \omega_\mu \,.\label{bscd}
\end{equation}

In the conformal gauge the metric 
$G_{\mu\nu}=e^a_\mu \star e^b_\nu \eta_{ab}$
and the volume two-form 
$E_{\mu\nu}=e^a_\mu \star e^b_\nu \varepsilon_{ab}$ are real.

\subsection{Dirac and Laplace operators}\label{sDL}
Next we like to define noncommutative
deformations of the Dirac and Laplace
operators. We require that these operators coincide with their
commutative counterparts in the limit $\theta\to 0$, and that
they are hermitian with respect to
some ``natural'' inner product. This procedure is
rather unrigorous, but, as we shall see below, the choice of
meaningful deformations is indeed very limited. 

We start with
the Dirac operator. Let $\gamma^\mu$ be the {\em flat} space
constant $\gamma$-matrices: 
\begin{equation}
\gamma^\mu \gamma^\nu +
\gamma^\nu \gamma^\mu = 2\delta^{\mu\nu} \,,
\label{defg}
\end{equation}
i.e. $(\gamma^1)^2=(\gamma^2)^2=1$. Let $\gamma_* = \gamma^1\gamma^2$.
This definition is convenient in the conformal gauge.

We have to fix a scalar product in the space of spinors. Let $\kappa_1$
and $\kappa_2$ be spinorial fields (this means simply two-component
complex fields in this context, nothing more). Then
\begin{equation}
\langle \kappa_1,\kappa_2 \rangle =\int d^2x \, \kappa_1^\dag \star
e^{2\rho}\star \kappa_2 \,.\label{spinpro}
\end{equation}
This product is linear, symmetric, and positive as long as both
$e^\rho$ and $e^{-\rho}$ are well defined. We can even get rid of
$e^{2\rho}$ by changing the variables to
the densitised fields $\tilde\kappa = e^{\rho} \kappa$. Then
\begin{equation}
\langle \kappa_1,\kappa_2 \rangle =
\langle \tilde \kappa_1,\tilde \kappa_2 \rangle_0=
\int d^2x \, \tilde \kappa_1^\dag \tilde \kappa_2 \,.\label{0npro}
\end{equation}
Note, that $\langle \ ,\ \rangle_0$ does not contain star at all.
By a similar change of the variables one can move $e^{2\alpha\rho}$,
with $\alpha$ being a real constant, in front of $\kappa_1^\dag$
in (\ref{spinpro}) so that the scalar product becomes:
\begin{equation}
\widetilde{\langle \kappa_1,\kappa_2 \rangle} =\int d^2x \, 
e^{2\alpha\rho}\star \kappa_1^\dag \star
e^{2(1-\alpha)\rho}\star \kappa_2 \,.\label{newnpro}
\end{equation}
This product seems to be the most general meaningful deformation
of the standard commutative inner product. 
We see, that although there exists different choices of the inner product,
they all are related by a change of variables.
We choose $\alpha=0$, i.e. the product defined in (\ref{spinpro}).

Let us define the Dirac operator by the equation
\begin{equation}
\Dir =i\gamma^\mu e^{-\rho}\star \left( \partial_\mu + 
\frac 12 \omega_\mu \gamma_* \right)\,. \label{Dir}
\end{equation}
This operator is fixed by its' commutative counterpart up to the order
in which we write $e^{-\rho}$ and $\omega$. This order is then uniquely 
defined by the requirement that $\Dir$ is hermitian\footnote{An operator
satisfying (\ref{hermD}) is also called symmetric or formally self-adjoined.
A self-adjoint operator must satisfy one additional requirement
regarding its domain. We shall not consider this requirement,
as well as we shall ignore such issues as completeness of the Hilbert
spaces etc.}
with respect to
the inner product (\ref{spinpro}):
\begin{equation}
\langle \Dir \star \kappa_1,\kappa_2 \rangle 
=\langle \kappa_1,\Dir \star \kappa_2 \rangle
\label{hermD}
\end{equation}
We stress that all multiplications are the Moyal star products. 
There is no additional ambiguity related to the choice of the Dirac 
operator. Let us remind that we have fixed $b_\mu=0$. Otherwise,
$b_\mu$ should have also appeared in (\ref{Dir}). 

Note, that the operator (\ref{Dir}) does not define any spectral triple
(cf. sec.\ 4.4 of \cite{Gayral:2004ww}).
The reason is that the commutator 
$[\Dir,f]=i\gamma^\mu [e^{-\rho},f] \partial_\mu + \dots$, 
where $f$ is a function, is not a bounded operator because of the presence
of the ``first-order'' term proportional to $\partial_\mu$.
This difficulty is hardly possible to avoid if one likes to identify
the leading symbol of the Dirac operator (i.e. the term appearing
in front of $\partial_\mu$) with a zweibein of a noncommutative
gravity theory on the Moyal plane (of which the noncommutative JT gravity
considered above is an example). 

By using the torsion constraint (\ref{ctors}) one can prove an
analog of the Lichnerowicz formula
\begin{equation}
\Dir^2=\Delta_{\mathrm{Spin}} + \frac 12 e^{-2\rho}\star \epsilon^{\mu\nu}
(\partial_\mu \omega_\nu ) + \frac 14 e^{-2\rho}\star \gamma_*
\epsilon^{\mu\nu} \omega_\mu \star \omega_\nu\,, \label{cLich}
\end{equation}
where the spinor Laplacian reads
\begin{equation}
\Delta_{\mathrm{Spin}}=-e^{-2\rho}\star \left(\partial_\mu 
+\frac 12 \gamma_* \omega_\mu \right)^2. \label{spinL}
\end{equation}
Note, that the 2nd and 3rd terms on the right hand side of (\ref{cLich})
are not necessarily real. The reason is that left multiplication by
a real function is not a hermitian operation with respect to the inner
product (\ref{spinpro}) if this function does not commute with $e^{2\rho}$.

Basing on (\ref{cLich}) we may conjecture that there exist 
generalisations of the Dirac and Laplace operator such that the
following formula holds for generic $e_\mu^a$ and generic
torsionless connection (including $b_\mu$):
\begin{equation}
\Dir^2=\Delta_{\mathrm{Spin}} +\frac 18 \{ e_a^\mu ,e_b^\nu \}\star
R_{\mu\nu}^{ab} +\frac 18 \{ e_a^\mu ,e_b^\nu \}\star \epsilon^{ab}
\eta_{cd} R_{\mu\nu}^{cd} \gamma_*\,. \label{Lich}
\end{equation}
The second term on the right hand side of (\ref{Lich}) is a rather
straightforward extension of the $R/4$ term appearing in commutative
theories\footnote{With the help of the identity
$\varepsilon^{ab}\varepsilon_{cd}=\delta^a_c \delta^b_d -
\delta^a_d \delta^b_c$ one can relate this term to the scalar curvature
density (\ref{scd}).}. 
The third term is a new feature of noncommutative theories.

Let us now turn to scalar fields. It is natural to assume that in the 
conformal gauge massless minimally coupled scalar fields decouple
from the geometry (as in the commutative case), so that the action
reads
\begin{equation}
S_{\mathrm{m}}=\frac 12 \int d^2x (\partial_\mu f^\dag )(\partial_\mu f)
\label{sact}
\end{equation}
Now we have to choose an inner product in the space of the scalar fields.
Let us fix it to be
\begin{equation}
\langle f_1,f_2 \rangle =\int d^2x\ f_1^\dag \star e^{2\rho} \star f_2
\label{sspro}
\end{equation}
in full analogy with the spinor case. We can change the variables,
$\tilde f=e^\rho f$, so that we obtain a ``trivial'' inner product
\begin{equation}
\langle\tilde f_1 ,\tilde f_2 \rangle_0=\int d^2x\ \tilde f_1^\dag 
\star \tilde f_2
=\int d^2x\ \tilde f_1^\dag \tilde f_2 \label{trin}
\end{equation} 
for the new fields $\tilde f$.Then 
\begin{equation}
S_{\mathrm{m}}=\frac 12 \int d^2x 
\tilde f^\dag \star \Delta \star \tilde f  \label{sLap}\,,
\end{equation}
where
\begin{equation}
\Delta = -e^{-\rho}\star \partial_\mu^2 e^{-\rho}. \label{Delta}
\end{equation}

Next we integrate (formally) over $\tilde f$ (cf. (\ref{Wm})) with the measure
$\mathcal{D}f^\alpha =\mathcal{D}\tilde f^\dag \mathcal{D}\tilde f$ 
to obtain the effective
action
\begin{equation}
W_{\mathrm{m}}= \ln \det\, \Delta \,.\label{sWm}
\end{equation} 
This expression is, of course, divergent and has to be regularised.
\subsection{Heat kernel and the anomaly}\label{shk}
Actual calculations of the effective action will be done in the
scalar case only (this case is already complicated enough).
We use the zeta function and heat kernel techniques
(see \cite{Vassilevich:2003xt} for a recent review)
to regularise the determinant (\ref{sWm}).
The heat kernel expansion on (flat) Moyal spaces was studied in
\cite{Vassilevich:2003yz,Gayral:2004ww}. 
We start with several definitions. Let $h$ be a smooth function, and
$t$ be a positive real number. Then the heat kernel (or, in a more precise
terminology,  the heat trace) is defined as
\begin{equation}
K(h,t,\Delta)=\mathrm{Tr}_{L^2} \left( h \star \exp (-t\Delta)\right),.
\label{defhk}
\end{equation}
The $L^2$ space is defined with respect to the inner product (\ref{trin}).
Here noncommutativity plays no role, and $L^2$ consists of all square
integrable functions on $\mathbb{R}^2$.
We shall assume that $\Delta$ is a positive operator, so that we can
define the zeta function:
\begin{equation}
\zeta (h,s,\Delta)=\mathrm{Tr}_{L^2} \left( h \star \Delta^{-s} \right)
\,.
\label{defzeta}
\end{equation}
The zeta function is related to the heat kernel by the integral
transformation
\begin{equation}
\zeta (h,s,\Delta)=\Gamma (s)^{-1} \int_0^\infty dt\, t^{s-1} K(h,t,\Delta)
\,.\label{zK}
\end{equation}
Note, that the existence of the heat kernel and of the zeta function
for the operator (\ref{Delta}) has not been rigorously stated so far
in the noncommutative case. Below we shall present some arguments
showing that these object do exist at the level of rigour
accepted in physics. We shall also demonstrate that, at least
in the sense of formal power series in $\rho$, there is an
asymptotic expansion as $t\to +0$:
\begin{equation}
K(h,t,\Delta)\simeq \sum_{k=0}^\infty t^{k-1} a_{2k} (h,\Delta)\,.
\label{hkas}
\end{equation}
As in the commutative case, there are no terms with fractional
powers of $t$. 

The heat kernel coefficients $a_{2k}$ are related to the residues
of $\zeta \Gamma$.
\begin{equation}
a_{2k}(h,\Delta)
=\mathrm{Res}_{s=1-k} \left( \Gamma (s) \zeta (h,s,\Delta)\right)\,.
\label{aRes}
\end{equation}
In particular,
\begin{equation}
a_2(h,\Delta)=\zeta (h,0,\Delta).\label{a2z0}
\end{equation}

It has been proposed in \cite{zeta}
to use the zeta function to regularise the effective
action:
\begin{equation}
W_s=-\mu^2 \int_0^\infty dt\, t^{s-1} K(t,\Delta)
= \mu^2 \Gamma (s) \zeta (s,\Delta) \,.\label{zreg}
\end{equation}
Here $\mu$ is a constant of the dimension of mass introduced
to keep proper dimension of the effective action. Spectral functions
without the first argument correspond to $h=1$, i.e. 
$K(t,\Delta):=K(1,t,\Delta)$, $\zeta (s,\Delta):=\zeta (1,s,\Delta)$.
The regularization is removed in the limit $s\to 0$. At $s=0$ the
effective action has a simple pole:
\begin{equation}
W_s=-\left ( \frac 1s -\gamma_E +\ln \mu^2 \right) \zeta (0,\Delta)
-\zeta'(0,\Delta)\,,\label{pole}
\end{equation}
where $\gamma_E$ is the Euler constant, prime denotes differentiation
with respect to $s$. The second term on the right hand side of
(\ref{pole}) is nothing else than the Ray-Singer definition of the
functional determinant \cite{RS71}.

The pole terms should be removed by the renormalization. The
remaining part of $W_s$ is the renormalised effective action:
\begin{equation}
W^{\mathrm{ren}}_{\mathrm{m}}=-\ln (\mu^2)\zeta (0,\Delta)
-\zeta'(0,\Delta)\,.\label{Wren}
\end{equation}
The constant $\mu^2$, which is arbitrary so far, describes the
renormalization ambiguity and has to be fixed by a normalisation
condition.

Note, that the operator $\Delta$ and, consequently, the
effective action (\ref{Wren}) depend on the conformal field
$\rho$ only. Therefore, we can use the conformal anomaly to
calculate $W^{\mathrm{ren}}_{\mathrm{m}}$. Let us rescale 
$\rho \to \alpha \rho$, where $\alpha$ is a real parameter
between $0$ and $1$. Then we introduce 
\begin{equation}
\Delta_{[\alpha]}=-e^{-\alpha\rho} \star \partial^2 e^{-\alpha\rho}
\label{Dal}
\end{equation}
and the corresponding effective action $W^{\mathrm{ren}}_{\mathrm{m}}(\alpha)$.
We may write
\begin{eqnarray}
&&\frac{d}{d\alpha} \zeta(s,\Delta_{[\alpha]})=
-s \mathrm{Tr} \left(\frac{d}{d\alpha} \Delta_{[\alpha]} \star
\Delta_{[\alpha]}^{-s-1} \right)=
2s \mathrm{Tr} \left( \rho \star \Delta_{[\alpha]}^{-s} \right) \nonumber \\
&&=2s \zeta (\rho, s,\Delta_{[\alpha]}) \,.\label{ddsz}
\end{eqnarray}
By combining (\ref{ddsz}) with (\ref{Wren}) and (\ref{a2z0}) we obtain:
\begin{equation}
\frac{d}{d\alpha} W^{\mathrm{ren}}_{\mathrm{m}}(\alpha) = -2 
a_2(\rho,\Delta_{[\alpha]}) .\label{convar}
\end{equation}
This quantity describes conformal non-invariance of the effective
action (conformal anomaly).

We arrive at the problem of calculating $a_2(h,\Delta)$. If we set
$h=1$ this calculation should give us the divergent part of the
effective action (\ref{pole}). For $h=\rho$ and $\rho \to \alpha\rho$
insider the operator the same calculation also defines the conformal
variation (\ref{convar}). We have to evaluate the small $t$ 
asymptotics of the heat trace for the noncommutative Laplacian
$\Delta$. Recently, it was realised \cite{Vassilevich:2003yz,Gayral:2004ww}
that the noncommutativity by itself is not a big problem. The problem
is that the star multiplication by a functions appears in a combination
with the highest derivatives of the operator (i.e. the star multiplication
enters the leading symbol of the operator). Heat trace calculations
with operators in which highest derivatives are multiplied by an
(almost) arbitrary matrix valued function appeared recently
in the context of a matrix generalisation of general relativity
\cite{Avramidi:2004fc}, and long ago in the context
of the bosonisation of QCD \cite{Andrianov:jh}. Our calculations (see
below) share some similarities with the above mentioned papers, though
our case is even more complicated.

The general strategy adopted here is taken from
\cite{Vassilevich:2003yz}. To evaluate (\ref{defhk}) we sandwich
the expression under the trace between plane waves and then integrate
over $x$ and $k$: 
\begin{equation}
K(h,t,\Delta)=\int d^2x \int \frac {d^2k}{(2\pi)^2} 
e^{-ikx} \star h(x) \star e^{-t\Delta} \star e^{ikx}\,. \label{Kii}
\end{equation}
Note, that the plane wave basis is orthormal with respect to the inner
product (\ref{trin}).
Next we push $e^{ikx}$ to the left to obtain:
\begin{eqnarray}
&&K(h,t,\Delta)=\int d^2x \int \frac {d^2k}{(2\pi)^2}
\, h(x) \nonumber \\
&&\qquad \qquad \qquad \quad \star \exp \left( -tk^2 e^{-2\rho} 
+te^{-\rho} \star \partial^2 e^{-\rho}
+2i e^{-\rho} \star (k\partial)e^{-\rho} \right),
\label{kij}
\end{eqnarray} 
where $(k\partial):=k^\mu \partial_\mu$, $k^2:=k^\mu k^\nu \delta_{\mu\nu}$.
The most important observation needed to derive (\ref{kij}) was made
in \cite{Vassilevich:2003yz}. Under the integral over $x$ the plane wave
$e^{ikx}$ can be pushed through the Moyal star without any modifications
of the latter\footnote{This means that in the derivation of this formula
the derivatives appearing in the Moyal product (\ref{starprod}) can be
safely ignored.}, 
so that the only effect of this operation is the replacement
$\partial^2\to (\partial +ik)^2$ in $\Delta$.  

Let $e^{-2\rho}$ be bounded from the below by a positive constant
$c$, $0<c\le e^{-2\rho}$. Then, for large $k$ the integrand in (\ref{kij})
falls off as $e^{-tck^2}$. Therefore, the integral over $k$ converges
for all $x$. For a sufficiently good localised function $h(x)$ the integral
over $x$ should also exist. This is the physical argument in favour
of the existence of the heat trace we have announced above. 

Actually, we are interested in the $t\to 0$ asymptotics of the heat
trace only. To evaluate these asymptotics one has to isolate
$\exp(-tk^2e^{-2\rho})$, expand the rest, and integrate over $k$.  
The following integrals will be useful,
\begin{eqnarray}
&&\int \frac {d^2k}{(2\pi)^2} e^{-ak^2} k^{2n}
=\frac {n!}{4\pi a^{n+1}} \,,\label{int1}\\
&&\int \frac {d^2k}{(2\pi)^2} e^{-ak^2}k_\mu k_\nu  k^{2n}
=\frac 12 \delta_{\mu\nu}
\frac {(n+1)!}{4\pi a^{n+2}} \,,\label{int2}
\end{eqnarray}
where $a=e^{-2\rho}t$, all exponentials and powers are defined with
the Moyal star product (for example, $a\star a^{-1}=1$). 
The formulae (\ref{int1}) and (\ref{int2}) are obvious in the commutative
case, but are less trivial in the noncommutative one. To obtain
(\ref{int1}) and (\ref{int2}) one has to represent $a=t+(e^{-2\rho}-1)t$,
keep $e^{-tk^2}$ and expand the rest into a power series in $\rho$.
Then one integrates over $k$ and sums up the series.

Before expanding the exponent in (\ref{kij}) it is useful to estimate
the power of $t$ in each of the individual terms. In terms of the
rescaled variable $\tilde k=t^{1/2} k$ the expression in the exponent
(\ref{kij}) reads:
\begin{equation}
-\tilde k^2 e^{-2\rho} + t^{1/2} 2i e^{-\rho}\star (\tilde k \partial)
e^{-\rho} + t e^{-\rho}\star \partial^2
e^{-\rho} \,.\label{resc}
\end{equation}
The integration measure $d^2k=t^{-1} d^2\tilde k$ produces an overall
factor of $t^{-1}$. The second term in (\ref{resc}) contributes $t^{1/2}$,
and the third contributes $t$ to the expansion. It is also clear that
all half-integer powers of $t$ vanish after integration over the momenta.

The coefficient $a_0(h,\Delta)$ is relatively easy to obtain:
\begin{equation}
a_0(h,\Delta)=\int d^2x h \star \int \frac {d^2\tilde k}{(2\pi)^2}
\exp (-e^{-2\rho}\tilde k^2)=\frac 1{4\pi} \int d^2x h \star e^{2\rho}\,.
\label{a0}
\end{equation}
This coefficient is an obvious generalisation of corresponding
commutative expression. 

Calculations of $a_2$ are much more involved. One has to keep the terms
which are either second order in the second term in (\ref{resc}),
or linear in the third term. The first term should be taken into account
exactly. After long but rather elementary calculations we obtain:
\begin{eqnarray}
&&a_2(h,\Delta)= \frac 1{4\pi} \int d^2x\, h(x) \nonumber \\
&&\qquad \star \left( \sum_{n=0}^\infty
\frac 1{n+1} e^{2(n+1)\rho} \star
\underbrace{ [e^{-2\rho},[e^{-2\rho},[\dots,
e^{-\rho}\star \partial^2 e^{-\rho}]]]}_{\mbox{$n$\ commutators}}- \right.
\label{a2ser}\\
&&\qquad -\sum_{m,n=0}^\infty \frac{2(n+m+1)! e^{2(n+m+2)\rho} }{
n!(m+1)! (n+m+2) }\star 
\underbrace{ [e^{-2\rho},[e^{-2\rho},[\dots,
e^{-\rho}\star \partial_\mu e^{-\rho}]]]}_{\mbox{$n$\ commutators}}
\nonumber \\
&&\qquad \qquad \qquad \qquad \qquad \star \left.
\underbrace{ [e^{-2\rho},[e^{-2\rho},[\dots,
e^{-\rho}\star \partial_\mu e^{-\rho}]]]}_{\mbox{$m$\ commutators}}
\right)\nonumber
\end{eqnarray}

It seems that the best one can do with (\ref{a2ser}) is to extract
several leading terms of an expansion in $\rho$. Fortunately, only
a finite number of terms in (\ref{a2ser}) contribute to any finite
order of this expansion. 
\begin{eqnarray}
&&a_2(h,\Delta)= \frac 1{4\pi} \int d^2x\, h(x)\star
\left( -\frac 13 \partial^2 \rho + \frac 1{30}
[[\rho,(\partial_\mu \rho)],(\partial_\mu \rho)]
\right.\nonumber\\
&&\qquad\qquad\qquad\qquad \left.
+\frac 7{90} \partial_\mu [[(\partial_\mu \rho),\rho],\rho] +
\mathcal{O} (\rho^4) \right)\,. \label{a2r3}
\end{eqnarray}

One may expect that $a_2$ is given by an analog of the ``commutative''
expression:
\begin{equation}
a_2(h,\Delta)_{\mathrm{com}} = \frac 1{4\pi} \int d^2x \frac 16
h(x) \mathcal{R} \,,\label{coma2}
\end{equation}
where $\mathcal{R}$ is the scalar curvature density (\ref{scd}).
With the help of (\ref{bscd}) and (\ref{expom}) one can easily check
that (\ref{coma2}) remains true in the noncommutative case in the
orders $\rho$ and $\rho^2$, but is violated in the order $\rho^3$.
The origin of this deviation is yet unclear.

Next we notice that $a_2(1,\Delta)$ is given by surface terms at
least up to the order $\rho^3$. Therefore, there are no local
divergences in (\ref{pole}), and no local counterterms are needed
for the renormalization.

Now we can integrate (\ref{convar}) to obtain an analog of the
Polyakov action \cite{Polyakov:1981rd} (see also \cite{Luscher:1980fr}).
It is natural to assume that $W_{\mathrm{m}}^{\mathrm{ren}}(\alpha=0)=0$.
Then
\begin{equation}
W_{\mathrm{m}}^{\mathrm{ren}}=-\frac 1{4\pi} \int d^2x \left(
-\frac 13 \rho\star \partial_\mu^2 \rho + \frac 1{45}
[\rho,\partial_\mu \rho]\star [\rho,\partial_\mu \rho] +
\mathcal{O}(\rho^5) \right)\,.\label{Polact}
\end{equation}
The first term on the right hand side is just the standard anomaly
induced action. The second term appears due to the noncommutativity and
vanishes in the commutative limit $\theta\to 0$.

We have to admit that our choice for the Laplacian is not unique.
There can be other NC Laplacians which may be more relevant for
physical or mathematical applications. However, main technical
tools developed in this paper should remain applicable for analysing
the heat kernel expansion for that other Laplacians, perhaps at the
expense of some technical complications.
\section{Conclusions}\label{scon}
In this paper we have studied quantisation of noncommutative gravity
theories in two dimensions. We started with the path integral in the
noncommutative JT model and demonstrated that all gravitational
degrees of freedom can be integrated out exactly even in the presence
of matter fields (with some minor restrictions on the interaction
between matter and gravity). The resulting quantum effective action 
coincides with the classical action for gravity plus an effective
action for the matter calculated as if the gravity fields
were classical. Then we studied the matter effective action
(restricting ourselves to the conformal gauge for simplicity).
We have found several natural differential operators which may
describe quantum matter fluctuations and studied their properties.
In particular, a noncommutative analog of the Lichnerowicz formula
has been derived. Then we turned to the scalar Laplacian,
evaluated two leading terms of the heat kernel expansion,
calculated conformal anomaly (as an expansion in the conformal
field $\rho$), and found a noncommutative analog of the Polyakov
action.

Note that the noncommutative JT gravity is the first noncommutative
(Moyal) quantum gravity model there one can go this far. Our main
message is, therefore, that noncommutative gravities can indeed
be successfully quantised. 

The results obtained here can be improved in many respects. Let
us outline just a few directions of possible future developments.
\begin{itemize}
\item It is interesting to construct noncommutative deformations
of two-dimen\-sional dilaton gravities other than JT. In this
more general context 
the gauge theory formulation (\ref{gact}) is not applicable.
\item In the matter sector we have only constructed a natural
fluctuation operator in the conformal gauge. It is important
to check whether this operator corresponds to an action which
possesses all necessary symmetries. With such an action at hand,
one may study many interesting phenomena as the black hole
formation, scattering on black holes (perhaps, with bound state formation
\cite {Vassilevich:2003he}), etc.
\item One definitely has to pay more attention to a careful
definition of the path integral, especially in the presence
of the space-time noncommutativity (cf.\ recent work 
\cite{Fujikawa}).
\item One has to undertake a more rigorous study of the heat
kernel and of the zeta function for the operators
on ``curved'' Moyal plane (or Moyal torus).
\item Although the noncommutative JT model \cite{Cacciatori:2002ib}
has a proper number of gauge symmetries, including Lorentz boosts and
diffeomorphisms, the restriction to a constant noncommutativity
parameter $\theta$ does not look very natural since it implicitly
selects a coordinate system. It is an interesting
problem to construct a gravity theory with the Kontsevich star 
\cite{Konstar} instead of the Moyal one.
\end{itemize}
\section*{Acknowledgements}
I am grateful to Daniel Grumiller and Max Kreuzer for many helpful discussions.
This work has been supported in part 
by the DFG project BO 1112/12-1.
\appendix
\section{Notations and conventions}\label{appnot}
Our sign conventions are mostly taken from \cite{Grumiller:2002nm}.
We use the tensor $\eta^{ab}=\eta_{ab}={\mathrm{diag}}(+1,-1)$ to move
indices up and down. The Levi-Civita tensor is defined by 
$\varepsilon^{01}=-1$, so that the following relations hold
\begin{equation}
\varepsilon^{10}=\varepsilon_{01}=1,\qquad
{\varepsilon^0}_1={\varepsilon^1}_0=-{\varepsilon_0}^1
=-{\varepsilon_1}^0=1\,.\label{epsrel}
\end{equation}
These relations are valid for both $\varepsilon^{ab}$ and 
$\varepsilon^{\mu\nu}$. Note, that $\varepsilon^{\mu\nu}$ is
always used with both indices up. 

We also use the light-cone basis in which
\begin{eqnarray}
&&\eta_{+-}=\eta_{-+}=\eta^{+-}=1,\qquad \eta^{++}=\eta_{++}=\eta^{--}
=\eta_{--}=0\,,\nonumber \\
&&{\varepsilon^+}_+=-{\varepsilon^-}_-=1,\qquad
\varepsilon_{-+}=-\varepsilon_{+-}=\varepsilon^{+-}=-\varepsilon^{-+}=1.
\label{lcsigns}
\end{eqnarray}

In sec.\ \ref{scona} we use the Euclidean signature, so that
\begin{equation}
\eta^{11}=\eta^{22}=1,\qquad \varepsilon^{12}=-\varepsilon^{21}=1\,.
\label{Euconv}
\end{equation}
\section{Symmetry transformations}\label{appsym}
Here we present an explicit form of the gauge symmetries of the
non-commutative JT model \cite{Cacciatori:2002ib}\footnote{Here we
correct a few misprints in corresponding formulae
of Ref.\ \cite{Cacciatori:2002ib}. The present author is grateful to
Sergio Cacciatori and Luca Martucci for correspondence regarding this
point.}.

Translations:
\begin{eqnarray}
&&\delta_\alpha e^a_\mu =\partial_\mu \alpha^a 
+ \frac 12 {\varepsilon^a}_b\{ \omega_\mu,\alpha^b\} 
+ \frac i2 [b_\mu,\alpha^a] \,,\nonumber\\
&&\delta_\alpha \phi^a=-\Lambda \left( 
\frac 12 {\varepsilon^a}_b \{ \phi,\alpha^b\}
+\frac i2 [\psi ,\alpha^a] \right) \,,\label{trans}\\
&&\delta_\alpha \omega_\mu = \frac \Lambda{2} 
\varepsilon_{ab} \{ e_\mu^a,\alpha^b\}, \qquad
\delta_\alpha \phi = -\frac 12 \varepsilon_{ab}\{ \phi^a,\alpha^b\} ,
\nonumber \\
&&\delta_\alpha b_\mu =\frac{i\Lambda}2 \eta_{ab} [e_\mu^a,\alpha^b],\qquad
\delta_\alpha \psi= -\frac i2 \eta_{ab} [\phi^a,\alpha^b].
\nonumber
\end{eqnarray}
Boosts:
\begin{eqnarray}
&&\delta_\xi e_\mu^a =-\frac 12 {\varepsilon^a}_b\{e_\mu^b,\xi\},
\qquad \delta_\xi \phi^a =-\frac 12 {\varepsilon^a}_b\{\phi^b,\xi\},
\nonumber \\
&&\delta_\xi \omega_\mu =\partial_\mu \xi +\frac i2 [b_\mu,\xi],
\qquad \delta_\xi \phi=\frac i2 [\psi,\xi],\label{boosts}\\
&&\delta_\xi b_\mu =-\frac i2 [\omega_\mu,\xi],\qquad
\delta_\xi \psi =-\frac i2 [\phi,\xi].\nonumber 
\end{eqnarray}
$U(1)$ gauge symmetry:
\begin{eqnarray}
&&\delta_\chi e^a_\mu =\frac i2 [e_\mu^a,\chi],\qquad
\delta_\chi \phi^a =\frac i2 [\phi^a,\chi],\nonumber \\
&&\delta_\chi \omega_\mu =\frac i2 [\omega_\mu,\chi],\qquad
\delta_\chi \phi =\frac i2 [\phi,\chi],\label{U1gauge}\\
&&\delta_\chi b_\mu =\partial_\mu \chi +\frac i2 [b_\mu,\chi],\qquad
\delta_\chi \psi =\frac i2 [\psi,\chi].\nonumber
\end{eqnarray}


\begin{thebibliography}{99}
\bibitem{ncreviews}
Douglas M. R. and Nekrasov N. A.,
Noncommutative field theory,
Rev.\ Mod.\ Phys.\  {\bf 73} (2001) 977-1029
[arXiv:hep-th/0106048];

Szabo R. J.,
Quantum field theory on noncommutative spaces,
Phys.\ Rept.\  {\bf 378} (2003) 207-299
[arXiv:hep-th/0109162].

\bibitem{MaBur}
M.~Buric and J.~Madore, Noncommutative 2-dimensional models of gravity,
preprint ESI 1467 (Vienna, 2004), arXiv:hep-th/0406232.

\bibitem{Chamseddine:1996zu}
A.~H.~Chamseddine and A.~Connes,
The spectral action principle,
Commun.\ Math.\ Phys.\  {\bf 186} (1997) 731
[arXiv:hep-th/9606001].

\bibitem{Chamseddine:2003cw}
A.~H.~Chamseddine,
Noncommutative gravity,
Annales Henri Poincare {\bf 4S2} (2003) S881
[arXiv:hep-th/0301112].

\bibitem{Gayral:2004ww}
V.~Gayral and B.~Iochum,
The spectral action for Moyal planes,
arXiv:hep-th/0402147.

\bibitem{Chamseddine:2000zu}
A.~H.~Chamseddine,
Complexified gravity in noncommutative spaces,
Commun.\ Math.\ Phys.\  {\bf 218} (2001) 283
[arXiv:hep-th/0005222].

\bibitem{Moffat:2000gr}
J.~W.~Moffat,
Noncommutative quantum gravity,
Phys.\ Lett.\ B {\bf 491} (2000) 345
[arXiv:hep-th/0007181].

\bibitem{Cacciatori:2002ib}
S.~Cacciatori, A.~H.~Chamseddine, D.~Klemm, L.~Martucci, 
W.~A.~Sabra and D.~Zanon,
Noncommutative gravity in two dimensions,
Class.\ Quant.\ Grav.\  {\bf 19} (2002) 4029 
[arXiv:hep-th/0203038].

\bibitem{Banados:2001xw}
M.~Banados, O.~Chandia, N.~Grandi, F.~A.~Schaposnik and G.~A.~Silva,
Three-dimensional noncommutative gravity,
Phys.\ Rev.\ D {\bf 64} (2001) 084012
[arXiv:hep-th/0104264].

\bibitem{Cardella:2002pb}
A.~H.~Chamseddine,
Invariant actions for noncommutative gravity,
J.\ Math.\ Phys.\  {\bf 44} (2003) 2534
[arXiv:hep-th/0202137];

M.~A.~Cardella and D.~Zanon,
Noncommutative deformation of four dimensional Einstein gravity,
Class.\ Quant.\ Grav.\  {\bf 20} (2003) L95
[arXiv:hep-th/0212071].

\bibitem{Obregon}
H.~Garcia-Compean, O.~Obregon, C.~Ramirez and M.~Sabido,
Noncommutative self-dual gravity,
Phys.\ Rev.\ D {\bf 68} (2003) 044015
[arXiv:hep-th/0302180];

H.~Garcia-Compean, O.~Obregon and C.~Ramirez,
Noncommutative topological half-flat gravity,
arXiv:hep-th/0402168.

\bibitem{Moffat}
J.~W.~Moffat,
Perturbative noncommutative quantum gravity,
Phys.\ Lett.\ B {\bf 493} (2000) 142
[arXiv:hep-th/0008089].

\bibitem{Quevedo}
H.~Quevedo and J.~G.~Tafoya,
Towards the deformation quantization of linearized gravity,
arXiv:gr-qc/0401088.

\bibitem{Ardalan:2002qt}
F.~Ardalan, H.~Arfaei, M.~R.~Garousi and A.~Ghodsi,
Gravity on noncommutative D-branes,
Int.\ J.\ Mod.\ Phys.\ A {\bf 18} (2003) 1051
[arXiv:hep-th/0204117].

\bibitem{Grumiller:2002nm}
D.~Grumiller, W.~Kummer and D.~V.~Vassilevich,
Dilaton gravity in two dimensions,
Phys.\ Rept.\  {\bf 369} (2002) 327
[arXiv:hep-th/0204253].

\bibitem{Kummer:1996hy}
W.~Kummer, H.~Liebl and D.~V.~Vassilevich,
Exact path integral quantization of generic 2-D dilaton gravity,
Nucl.\ Phys.\ B {\bf 493} (1997) 491
[arXiv:gr-qc/9612012].

\bibitem{JT}
C.~Teitelboim,
Gravitation And Hamiltonian Structure In Two Space-Time Dimensions,
Phys.\ Lett.\ B {\bf 126} (1983) 41;
The Hamiltonian Structure of Two-Dimensional Space-Time and its Relation
with the Conformal Anomaly,
in Quantum Theory Of Gravity, p. 327-344, S.Christensen (ed.), Adam Hilgar,
Bristol;

R.~Jackiw,
Liouville Field Theory: A Two-Dimensional Model For Gravity?
in Quantum Theory Of Gravity, p. 403-420, S.Christensen (ed.), Adam Hilgar,
Bristol; 
Lower Dimensional Gravity,
Nucl.\ Phys.\ B {\bf 252} (1985) 343.

\bibitem{Polyakov:1981rd}
A.~M. Polyakov, Quantum geometry of bosonic strings, Phys. Lett. {\bf
  B103} (1981)
207--210.

\bibitem{BerezinShubin}
F.~A.~Berezin and M.~A.~ Shubin,
{\it The Schr\"{o}dinger Equation},
Kluwer, Dordrecht, 1991.


\bibitem{Bayen:1977ha}
F.~Bayen, M.~Flato, C.~Fronsdal, A.~Lichnerowicz and D.~Sternheimer,
Deformation Theory And Quantization. 1. 
Deformations Of Symplectic Structures,
Annals Phys.\  {\bf 111} (1978) 61-110.


\bibitem{Kummer:1997jj}
W.~Kummer, H.~Liebl and D.~V.~Vassilevich,
Non-perturbative path integral of 2d dilaton gravity and two-loop  effects
from scalar matter,
Nucl.\ Phys.\ B {\bf 513} (1998) 723
[arXiv:hep-th/9707115].

\bibitem{Kummer:1998zs}
W.~Kummer, H.~Liebl and D.~V.~Vassilevich,
Integrating geometry in general 2D dilaton gravity with matter,
Nucl.\ Phys.\ B {\bf 544} (1999) 403
[arXiv:hep-th/9809168].

\bibitem{Bergamin:2004us}
L.~Bergamin, D.~Grumiller and W.~Kummer,
Quantization of 2D dilaton supergravity with matter,
arXiv:hep-th/0404004.

\bibitem{Grumiller:2003mc}
D.~Grumiller, W.~Kummer and D.~V.~Vassilevich,
Positive specific heat of the quantum corrected dilaton black hole,
J. High Energy Phys. {\bf 0307} (2003) 009
[arXiv:hep-th/0305036].

\bibitem{Faddeev:fc}
L.~D.~Faddeev and V.~N.~Popov,
Feynman Diagrams For The Yang-Mills Field,
Phys.\ Lett.\ B {\bf 25} (1967) 29.

\bibitem{Barnich:2003wq}
G.~Barnich, F.~Brandt and M.~Grigoriev,
Local BRST cohomology and Seiberg-Witten maps in noncommutative Yang-Mills
theory,
Nucl.\ Phys.\ B {\bf 677} (2004) 503
[arXiv:hep-th/0308092].

\bibitem{Jackiw:2001jb}
R.~Jackiw and S.~Y.~Pi,
Covariant coordinate transformations on noncommutative space,
Phys.\ Rev.\ Lett.\  {\bf 88} (2002) 111603
[arXiv:hep-th/0111122].

\bibitem{strip}
A.~H.~Chamseddine, G.~Felder and J.~Fr\"{o}hlich,
Gravity in noncommutative geometry,
Commun.\ Math.\ Phys.\  {\bf 155} (1993) 205
[arXiv:hep-th/9209044];

A.~H.~Chamseddine, J.~Fr\"{o}hlich and O.~Grandjean,
The Gravitational sector in the Connes-Lott formulation of the standard
model,
J.\ Math.\ Phys.\  {\bf 36} (1995) 6255
[arXiv:hep-th/9503093];

A.~Connes,
Noncommutative Geometry And Reality,
J.\ Math.\ Phys.\  {\bf 36} (1995) 6194.

\bibitem{Vassilevich:2003xt}
D.~V.~Vassilevich,
Heat kernel expansion: User's manual,
Phys.\ Rept.\  {\bf 388} (2003) 279
[arXiv:hep-th/0306138].

\bibitem{Vassilevich:2003yz}
D.~V.~Vassilevich,
Non-commutative heat kernel, Lett.\ Math.\ Phys.\ {\bf 67} (2004) 185
[arXiv:hep-th/0310144].

\bibitem{zeta}
J.~S.~Dowker and R.~Critchley,
Effective Lagrangian And Energy Momentum Tensor In De Sitter Space,
Phys.\ Rev.\ D {\bf 13} (1976) 3224;

S.~W.~Hawking,
Zeta Function Regularization Of Path Integrals In Curved Space-Time,
Commun.\ Math.\ Phys.\  {\bf 55} (1977) 133.

\bibitem{RS71}
D.~Ray and I.~M.~Singer,
R-torsion and the Laplacian on Riemannian manifolds,
Adv. Math. {\bf 7} (1971) 145-210.

\bibitem{Avramidi:2004fc}
I.~G.~Avramidi,
Gauged Gravity via Spectral Asymptotics of non-Laplace type Operators,
JHEP {\bf 0407} (2004) 030 [arXiv:hep-th/0406026].

\bibitem{Andrianov:jh}
A.~A.~Andrianov, V.~A.~Andrianov, V.~Y.~Novozhilov and Y.~V.~Novozhilov,
Joint Chiral And Conformal Bosonization In QCD And The Linear Sigma Model,
Phys.\ Lett.\ B {\bf 186} (1987) 401.


\bibitem{Luscher:1980fr}
M.~L{\"{u}}scher, K.~Symanzik, and P.~Weisz, Anomalies of the free loop wave
  equation in the {WKB} approximation, Nucl. Phys. {\bf B173} (1980)
365.

\bibitem{Vassilevich:2003he}
D.~V.~Vassilevich and A.~Yurov,
Space-time non-commutativity tends to create bound states,
Phys.\ Rev.\ D {\bf 69} (2004) 105006
[arXiv:hep-th/0311214].

\bibitem{Fujikawa}
K.~Fujikawa,
Path Integral for Space-time Noncommutative Field Theory,
Phys.\ Rev.\ D {\bf 70} (2004) 085006 [arXiv:hep-th/0406128].

\bibitem{Konstar}
M.~Kontsevich,
Deformation quantization of Poisson manifolds, I,
Lett.\ Math.\ Phys.\  {\bf 66} (2003) 157
[arXiv:q-alg/9709040];

A.~S.~Cattaneo and G.~Felder,
On the globalization of Kontsevich's star product and the perturbative
Poisson sigma model,
Prog.\ Theor.\ Phys.\ Suppl.\  {\bf 144} (2001) 38
[arXiv:hep-th/0111028].
\end{thebibliography}
\end{document}